\documentclass[5p]{elsarticle}
\journal{Physics Letters A}

\usepackage{amssymb}
\usepackage{xcolor} 
\usepackage{mathtools} 
\usepackage{layouts} 
\usepackage{scrextend}
\usepackage{ulem}

\graphicspath{{./figures/}}
\usepackage{tikz} 
\usetikzlibrary{arrows.meta,decorations.pathmorphing,shadows,positioning}

\usepackage{xifthen} 
\usepackage{mleftright} \mleftright 
\usepackage{microtype} 
\usepackage{xspace} 

\usepackage[psdextra]{hyperref}

\newcommand{\abs}[1]{\left| #1 \right|}
\newcommand{\br}[1]{\left( #1 \right)}
\newcommand{\brr}[1]{\left[ #1 \right]}
\newcommand{\brrr}[1]{\left\{ #1 \right\}}

\newcommand{\e}{\mathrm{e}} 
\newcommand{\I}{\mathrm{i}} 
\newcommand{\vect}[1]{\boldsymbol{#1}} 

\DeclareMathOperator*{\Tr}{Tr} 

\newcommand{\ket}[1]{\vert #1 \rangle} 
\newcommand{\braket}[2]{\langle #1 | #2 \rangle} 
\newcommand{\ketbra}[3][]{\left| #2 \right>
  \ifthenelse{\isempty{#1}}{\!}{_{\!#1}} \left< #3 \right|} 
\newcommand{\de}{\mathrm{d}}
\newcommand{\pd}[2]{\frac{\partial #1}{\partial #2}} 
\renewcommand{\wp}{\omega_\mathrm{p}}
\newcommand{\ws}{\omega_\mathrm{s}}
\newcommand{\Op}{\Omega_\mathrm{p}}
\newcommand{\Os}{\Omega_\mathrm{s}}
\newcommand{\Oo}{\Omega_0}
\newcommand{\Om}{\Omega_\mathrm{max}}
\newcommand{\ti}{t_\mathrm{i}}
\newcommand{\tf}{t_\mathrm{f}}
\newcommand{\Eg}{\omega_\mathrm{g}}
\newcommand{\Ee}{\omega_\mathrm{e}}
\newcommand{\Er}{\omega_\mathrm{r}}
\newcommand{\Dp}{\Delta_\mathrm{p}}
\newcommand{\D}{\Delta_3}
\newcommand{\Lio}{\mathcal{L}}
\newcommand{\U}{R(t,\vect{u}(t))}
\newcommand{\F}{\mathcal{F}}
\newcommand{\ak}{\vect{\alpha}_k}

\newcommand*{\eg}{e.g.\@\xspace}
\newcommand*{\ie}{i.e.\@\xspace}

\begin{document}
\begin{frontmatter}
  \title{A Tutorial on Optimal Control and Reinforcement Learning methods for
    Quantum Technologies\tnoteref{t1}}

  \tnotetext[t1]{CEWQOnline Special Issue.}

  \author[dfa,cnrimm]{Luigi Giannelli\corref{cor1}}
  \ead{luigi.giannelli@dfa.unict.it}
  \author[belfast]{Pierpaolo Sgroi}
  \author[belfast]{Jonathon Brown}
  \author[aalto]{Gheorghe Sorin Paraoanu}
  \author[belfast]{Mauro Paternostro}
  \author[dfa,infnct,cnrimm]{Elisabetta Paladino}
  \author[dfa,infnct,cnrimm]{Giuseppe Falci}

  \address[dfa]{Dipartimento di Fisica e Astronomia ``Ettore Majorana",
    Università di Catania, Via S. Sofia 64, 95123, Catania, Italy}
  \address[cnrimm]{CNR-IMM, UoS Università, 95123, Catania, Italy}
  \address[infnct]{INFN, Sez. Catania, 95123, Catania, Italy}
  \address[belfast]{Centre for Theoretical Atomic, Molecular, and Optical
    Physics, School of Mathematics and Physics, Queens University, Belfast BT7
    1NN, United Kingdom}
  \address[aalto]{QTF Centre of Excellence, Department of Applied Physics, Aalto
    University School of Science, P.O. Box 15100, FI-00076 AALTO, Finland}

  \cortext[cor1]{Corresponding author}

  \begin{abstract}
    Quantum Optimal Control is an established field of research which is
    necessary for the development of Quantum Technologies. In recent years,
    Machine Learning techniques have been proved usefull to tackle a variety of
    quantum problems. In particular, Reinforcement Learning has been employed to
    address typical problems of control of quantum systems. In this tutorial we
    introduce the methods of Quantum Optimal Control and Reinforcement Learning
    by applying them to the problem of three-level population transfer. The
    \textit{jupyter notebooks} to reproduce some of our results are open-sourced
    and available on \textit{github}\footref{fn:githublink}.

\end{abstract}

  \begin{keyword}
    quantum technologies \sep quantum control \sep optimal control \sep machine
    learning \sep reinforcement learning \sep STIRAP
  \end{keyword}
\end{frontmatter}



\section{\label{sec:introduction}Introduction}
In the last two decades many advances have been made in the field of
\textit{Quantum Technology} (QT). QT aims at developing practical applications
by making use of the properties of quantum mechanics, such as superposition and
entanglement~\cite{AcinNJP2018quantum}. The ability to precisely manipulate
quantum systems is a key tool in developing quantum
technologies~\cite{AcinNJP2018quantum,GlaserEPJD2015training,BoscainPRXQ2021introduction,RemboldAQS2020introduction,WilhelmAQ2020introduction}.

\textit{Quantum Control} (QC) looks at providing the user with a set of
time-dependent control parameters in order to drive a dynamical quantum system
such that it performs some specific
task~\cite{GlaserEPJD2015training,BoscainPRXQ2021introduction,RemboldAQS2020introduction,WilhelmAQ2020introduction}.

\textit{Optimal Control Theory} (OCT) is a field of applied mathematics and is
a powerful tool that provides methods to find controls that allow a dynamical
system to evolve to achieve a predefined goal. For reference textbooks see for
example~\cite{Liberzon2012calculus,dAlessandro2021introduction,Bressan2007introduction}.
When this theory is applied to quantum systems it is often referred to as
\textit{Quantum Optimal Control}
(QOC)~\cite{GlaserEPJD2015training,BoscainPRXQ2021introduction,RemboldAQS2020introduction,WilhelmAQ2020introduction}.
 However there are other techniques such as \textit{Transitionless Quantum
  Driving}~\cite{BerryJPAMT2009transitionless} (or \textit{Shortcut to
  Adiabaticity}~\cite{Guery-OdelinRMP2019shortcuts}).

More recently the overlap between the fields of machine learning and quantum
mechanics have been explored
extensively~\cite{DunjkoRPP2018machine,CarleoRMP2019machine,MehtaPR2019highbias,MarquardtSPLN2021machine,AlchieriQMI2021introduction},
with both machine learning algorithms used to improve the understanding and the
control of quantum
systems~\cite{CarleoS2017solving,YoussrynQI2020characterization,SgroiPRL2021reinforcement},
and properties of quantum mechanics used to improve machine learning algorithms
(Quantum Machine
Learning)~\cite{BeerNC2020training,RomeroQST2017quantum,SaggioN2021experimental}.
In particular, reinforcement learning (RL) has been employed in the context of
control of multi-level
systems~\cite{BrownNJP2021reinforcement,PorottiCP2019coherent,PaparellePLA2020digitally},
and for quantum sensing and
metrology~\cite{CostaEQT2021benchmarking,HentschelPRL2010machine,HentschelPRL2011efficient}.

In this tutorial we illustrate the methods of numerical Optimal Control and
Reinforcement Learning by applying them to the problem of population transfer in
a three-level $\Lambda$ or ladder system, for which a well-known solution is
STIRAP~\cite{KuklinskiPRA1989adiabatic,BergmannRMP1998coherent,VitanovRMP2017stimulated}.
Analytical solutions to this problem via Optimal Control have been presented
in~\cite{BoscainJMP2002optimal,YuanPRA2012controllability,BoscainPRXQ2021introduction}
and a numerical example is given in~\cite{GoerzSP2019krotov,optimization}, but,
to our knowledge, no systematic numerical study exists. This problem has also
been studied by applying Reinforcement Learning in
\cite{PorottiCP2019coherent,PaparellePLA2020digitally}. Here we improve those
results by defining the Markov Decision Process in a more convenient way, using
a simpler Reinforcement Learning algorithm, reaching the solution by training
the model for fewer episodes and reaching an overall better fidelity.

The paper is organized as follows: in Sec.~\ref{sec:three-level-popul} we
introduce the problem of three-level population transfer and its solution via
STIRAP. In Sec.~\ref{sec:optimal-control} we introduce Optimal Control Theory
and show its application on three-level population transfer. In
Sec.~\ref{sec:reinf-learn} we introduce the Reinforcement Learning paradigm and
show its application to the same problem. Finally, in Sec.~\ref{sec:conclusions}
we draw the conclusions.

We
open-source\footnote{\label{fn:githublink}\url{https://www.github.com/luigiannelli/threeLS_populationTransfer}}
parts of the source code we produced. It is easily adaptable, with minor
changes, to different situations involving population transfer in three-level
systems.

\section{Three-level population transfer\label{sec:three-level-popul}}
\begin{figure}[!ht]
  \centering
  \begin{tikzpicture}

  \tikzset{>=latex}

  \tikzset{wavy/.style={decorate, decoration = {snake,pre length=3pt,post length=7pt,}}}

  \def\sL{1.7}
  \def\spacing{0.7}
  \def\Ee{3}
  \def\Er{0.7}
  \def\detp{0.5}
  \def\det{0.4}
  \def\Eee{2.0}
  \def\Err{3.7}

  \node[font=\bfseries] at (0,\Err*0.97) {(a)};
  \node[font=\bfseries] at (2*\sL+2*\spacing,\Err*0.97) {(b)};






  
  \draw[thick] (0,0) node[left] {$\ket{g}$}-- ++ (\sL,0);
  \draw[thick] (\spacing,\Ee) node[left] {$\ket{e}$} -- ++ (\sL,0);
  \draw[thick] (\sL+\spacing,\Er) node[left] {$\ket{r}$} -- ++ (\sL,0);

  \draw[dotted] (\spacing,\Ee+\detp) -- ++ (\sL,0);
  \draw[dotted] (\sL+\spacing,\Er+\det) -- ++ (\sL,0);

  \draw[->,>=stealth, thick] (\spacing*0.9+\sL*0.9,\Ee) -- ++ (0,\detp)
  node[pos=0.4,right] {$\Dp$};

  \draw[->,>=stealth, thick] (\spacing*0.9+\sL*0.9*2.2,\Er) -- ++ (0,\det)
  node[pos=0.4,right] {$\D$};


  \draw[<->,color=blue, very thick] (\sL*0.5,0) -- (\spacing+\sL/2,\Ee+\detp)
  node[pos=0.5,left]{$\Op(t)$};
  
  \draw[<->,color=red, very thick] (\spacing+\sL/2,\Ee+\detp) --
  (\spacing+1.5*\sL, \Er+\det) node[pos=0.6,left]{$\Os(t)$};


  
  \draw[thick] (2*\sL+3.5*\spacing,0) node[left] {$\ket{g}$}-- ++ (\sL,0);
  \draw[thick] (2*\sL+3.5*\spacing,\Eee) node[left] {$\ket{e}$}-- ++ (\sL,0);
  \draw[thick] (2*\sL+3.5*\spacing,\Err) node[left] {$\ket{r}$}-- ++ (\sL,0);

  \draw[dotted] (2*\sL+3.5*\spacing,\Eee-\detp) -- ++ (\sL,0);
  \draw[dotted] (2*\sL+3.5*\spacing,\Err-\det) -- ++ (\sL,0);

  \draw[->,>=stealth, thick] (2.9*\sL+3.5*\spacing,\Eee) -- ++ (0,-\detp)
  node[pos=0.4,right] {$\Dp$};

  \draw[->,>=stealth, thick] (2.9*\sL+3.5*\spacing,\Err) -- ++ (0,-\det)
  node[pos=0.4,right] {$\D$};


  \draw[<->,color=blue, very thick] (2.5*\sL+3.5*\spacing,0) -- ++ (0,\Eee-\detp)
  node[pos=0.5,left]{$\Op(t)$};
  
  \draw[<->,color=red, very thick] (2.5*\sL+3.5*\spacing,\Eee-\detp) -- ++
  (0, \Err-\Eee-\det+\detp) node[pos=0.61,left]{$\Os(t)$};

\end{tikzpicture}
  \caption{\label{fig:threelevelsystem}Scheme of the three-level system. (a)
    $\Lambda$ structure and (b) ladder structure.}
\end{figure}
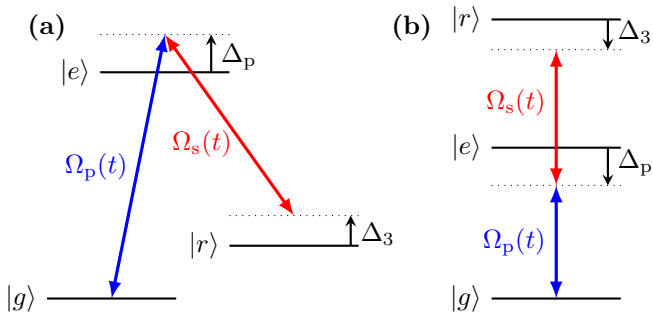

Consider a three-level system composed of the quantum states
$\{\ket{g}, \ket{e}, \ket{r}\}$ with energies
$\{\hbar\Eg, \hbar\Ee, \hbar\Er\}$. The transition $g\leftrightarrow e$ is
driven by a classical field (called \textit{pump} field) with frequency $\wp$
and Rabi frequency $\Op(t)$. The transition $e\leftrightarrow r$ is driven by
another classical field (called \textit{Stokes} field) at frequency $\ws$ and
Rabi frequency $\Os(t)$, see Fig.~\ref{fig:threelevelsystem}. Both Rabi
frequencies $\Op(t)$ and $\Os(t)$ can vary with time.

The process we want to achieve consists in transferring the population from
state $\ket{g}$ to state $\ket{r}$ by suitably shaping the Rabi frequencies
$\Op(t)$ and $\Os(t)$ in time.

In the following sections we introduce the equations that describe the dynamics
of the system, and the
STIRAP~\cite{KuklinskiPRA1989adiabatic,BergmannRMP1998coherent,VitanovRMP2017stimulated}
protocol which allows an efficient transfer of population.

\subsection{Master equation}
In the rotating wave approximation~\cite{RabiRMP1954use,Shore1990theory} and in
a convenient rotating frame the Hamiltonian
reads~\cite{BergmannRMP1998coherent}
\begin{equation}
  \label{eq:threelevelHamiltonian}
  \begin{aligned}
    \frac{H(t)}{\hbar} ={} &\Dp\ketbra{e}{e} + \D\ketbra{r}{r} + \frac{\Op(t)}{2}
    \br{\ketbra{g}{e} + \ketbra{e}{g}} +{} \\
    &+\frac{\Os(t)}{2} \br{\ketbra{e}{r} + \ketbra{r}{e}}
  \end{aligned}
\end{equation}
where the detunings from the resonances are defined as $\Dp=\wp-(\Ee-\Eg)$,
$\Delta_\mathrm{s}=\ws-\abs{\Er-\Ee}$, and $\D=\Dp-\Delta_\mathrm{s}$ for the
$\Lambda$ configuration, and $\D=\Dp+\Delta_\mathrm{s}$ for the ladder
configuration, see Fig.~\ref{fig:threelevelsystem}. While single-photon
resonance is not required in order to obtain a nearly perfect transfer, the
two-photon resonance is usually required~\cite{BergmannRMP1998coherent} (apart
some peculiar cases such as Ref.~\cite{SgroiPRL2021reinforcement}). Thus we
allow the single-photon detuning to be different from zero $\Dp\neq 0$, while
we assume the two-photon detuning to be zero $\D=0$ for the rest of this
manuscript. We also assume the Rabi frequencies $\Op(t)$ and $\Os(t)$ to be
real since their phase could be for example absorbed in the definition of the
states $\ket{g}$ and $\ket{r}$~\cite{Shore1990theory}. If the system under
consideration is an atomic or molecular system, then the Rabi frequencies are
given by $\Op(t) = -d_\mathrm{ge}\mathcal{E}_\mathrm{p}(t)/\hbar$ and
$\Os(t) = -d_\mathrm{er}\mathcal{E}_\mathrm{s}(t)/\hbar$, where $d_{mn}$ are
the components of the dipole-transition moments along their respective
electric-field vectors, and $\mathcal{E}_\mathrm{s/p}(t)$ are the slowly
varying amplitudes of the pump and Stokes electric
fields~\cite{VitanovRMP2017stimulated}.

In both the configurations $\Lambda$ and ladder (see
Fig.~\ref{fig:threelevelsystem}), the excited states can undergo spontaneous
emission to lower-lying states. Those emission processes that lead to levels
outside the three-level system determine a probability loss and thus are
undesirable. Spontaneous emission processes back to state $\ket{g}$ or
$\ket{r}$ are incoherent and thus are also undesirable.

In this work we only consider radiative decay from the excited state $\ket{e}$
to states outside the three-level system. We model this phenomenon by a
Born-Markov process described by the superoperator $\Lio_\gamma$ such that the
master equation describing the time evolution of the density matrix $\rho(t)$
reads~\cite{Carmichael1999statistical}
\begin{equation}
  \label{eq:mastereq}
  \dot \rho(t) = -\frac{\I}{\hbar} [H(t), \rho(t)] + \Lio_\gamma \rho(t),
\end{equation}
with
\begin{equation}
  \label{eq:liogamma}
  \Lio_\gamma \rho(t) =  \frac{\gamma}{2}
  \br{2\ketbra{s}{e}\rho(t)\ketbra{e}{s} - \ketbra{e}{e}\rho(t)
    - \rho(t)\ketbra{e}{e}}.
\end{equation}
Here $\ket{s}$ is an auxiliary state where the population losses at rate
$\gamma$ from state $\ket{e}$ are collected.

The figure of merit we use to quantify the performance of a protocol is the
\textit{fidelity} defined as
\begin{equation}
  \label{eq:fidelity}
  \F = \lim_{t\to\infty}\Tr\brrr{\rho(t)\ketbra{r}{r}}.
\end{equation}
It is clear that a perfect protocol would have fidelity $\F=1$.

\subsection{Review of the STIRAP protocol}
\textit{STImulated Raman Adiabatic Passage}
(STIRAP)~\cite{KuklinskiPRA1989adiabatic,BergmannRMP1998coherent,VitanovRMP2017stimulated}
is an adiabatic protocol that allows population transfer from state $\ket{g}$
to state $\ket{r}$ with fidelity close to one by keeping the population on the
lossy state $\ket{e}$ very low during the evolution. In order to explain the
STIRAP protocol we first introduce the \textit{adiabatic
  theorem}~\cite{BornZFP1928beweis,Messiah1961quantum}.

\subsubsection{Adiabatic theorem}
Given a time-dependent Hamiltonian $H_0(t)$, its instantaneous eigenstates
$\ket{n(t)}$ and its instantaneous eigenenergies $E_n(t)$ are given by
\begin{equation}
  \label{eq:instantaneoustimeindSchroedingerEq}
  H_0(t)\ket{n(t)} = E_n(t)\ket{n(t)},
\end{equation}
\ie, they are obtained by diagonalizing the Hamiltonian $H_0(t)$ at each time
step $t$. For simplicity lets assume all the states $\ket{n(t)}$ to be
non-degenerate for any $t$. The solution of the time-dependent Schr\"odinger
equation
\begin{equation}
  \label{eq:Schroedingereq}
  \I\hbar\pd{\ket{\psi(t)}}{t}=H_0(t)\ket{\psi(t)},
\end{equation}
is in general a linear combination of all the instantaneous eigenstates
\begin{equation}
  \label{eq:generalsolutionSchroedeq}
  \ket{\psi(t)} = \sum_n c_n(t) \ket{n(t)},
\end{equation}
where $c_n(t)$ are time-dependent complex amplitudes and
$\sum_n \abs{c_n(t)}^2=1$.

\begin{figure}[!ht]
  \centering
  \includegraphics[width=0.93\linewidth]{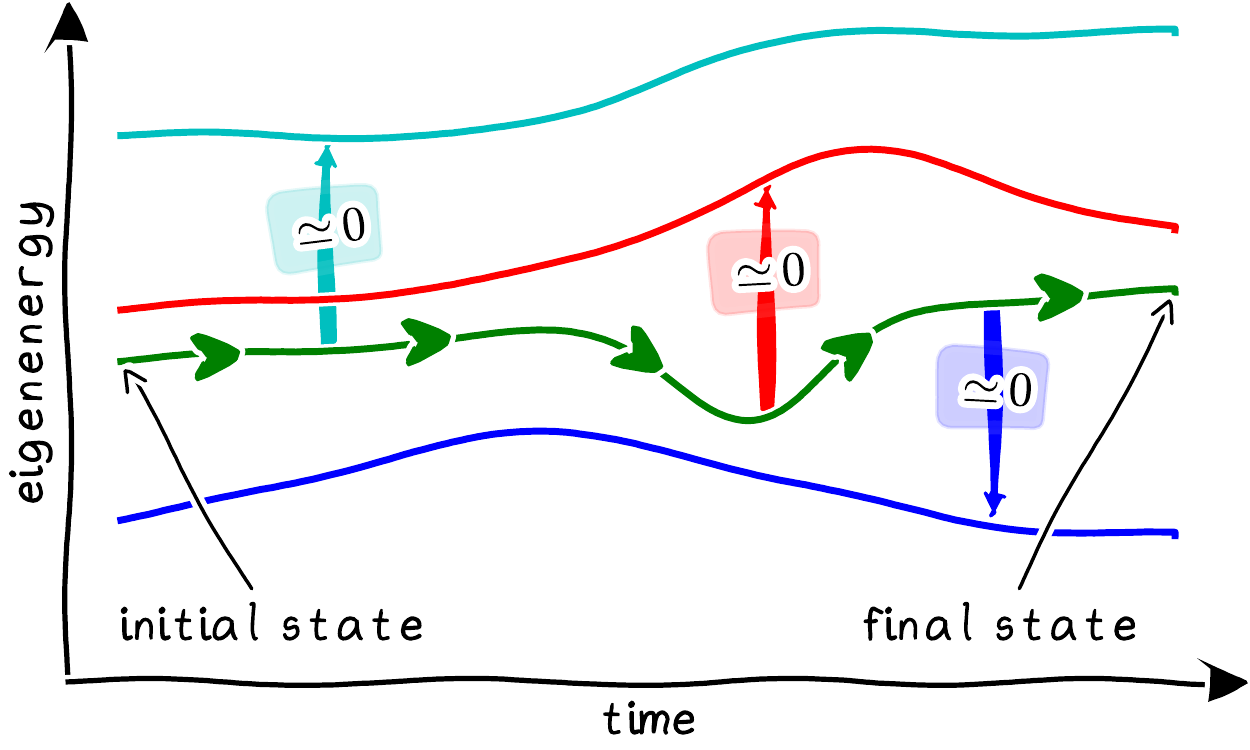}
  \caption{\label{fig:eigenfollow}Artistic view of adiabatic following of an
    instantaneous eigenstate. The colored boxes represent the transition
    probabilities which are approximately zero in an adiabatic process.}
\end{figure}

If the Hamiltonian $H_0(t)$ is \textit{slowly varying}\footnote{the meaning of
  \textit{slowly} is specified later in the text and summarized by
  eq.~\eqref{eq:adiabaticondition}.} and the initial state is one of the
instantaneous eigenstates, then the adiabatic theorem guarantees that the system
will follow that instantaneous eigenstate closely: during the time evolution of
the system, the transition amplitudes to instantaneous eigenstates different
from the starting one are approximately to zero (see Fig.~\ref{fig:eigenfollow}
for an \textit{artistic representation} of this concept). If the initial state
is $\ket{\psi(\ti)} = \ket{m(\ti)}$, then
\begin{equation}
  \label{eq:adiabaticevolution}
  \ket{\psi(t)}\simeq\e^{\I\alpha_m(t)}\ket{m(t)},
\end{equation}
\ie, $c_n(t)\simeq\e^{\I\alpha_m(t)}\delta_{mn}$, where $\alpha(t)$ is a
global phase\footnote{$\alpha(t)$ is the sum of the dynamic phase factor and
  the geometric phase.} which is not important for our discussion.

The condition for the Hamiltonian $H_0(t)$ to be considered slowly varying can
be obtained by imposing that the probability of finding the system in a state
$\ket{m(t)}$ different from the initial state $\ket{n(\ti)}$ is
small. This can be written as~\cite{Messiah1961quantum}
\begin{equation}
  \label{eq:adiabaticondition}
  \hbar\abs{\braket{n(t)}{\partial_t m(t)}}\ll \abs{E_n(t)-E_m(t)}, \forall
  m\neq n.
\end{equation}

We proceed by calculating the instantaneous eigenenergies and eigenstates (\ie,
the eigensystem) of Hamiltonian $H(t)$ reported in eq.~\eqref{eq:threelevelHamiltonian}.

\subsubsection{Eigensystem of $H(t)$}
The analysis of the three-level dynamics can be written in a simpler form by
defining
\begin{subequations}
  \begin{gather}
    \Oo(t) = \sqrt{\Op(t)^2 + \Os(t)^2}, \label{eq:effectiveRabifreq} \\
    \tan\theta(t) = \frac{\Op(t)}{\Os(t)}, \label{eq:tantheta} \\
    \tan\phi(t) = \frac{\Oo(t)}{\Dp + \sqrt{\Dp^2 + \Oo(t)^2}}.
    \label{eq:tanphi}
  \end{gather}
\end{subequations}
The instantaneous eigenvalues of Hamiltonian $H(t)$, eq.~\eqref{eq:threelevelHamiltonian}
(with $\D=0$) are~\cite{BergmannRMP1998coherent}
\begin{subequations}
  \label{eq:eigenvalues}
  \begin{gather}
    \lambda_0(t) = 0, \\
    \lambda_-(t) = -\frac{\hbar}{2}\Oo(t)\tan\phi(t), \\
    \lambda_+(t) = \frac{\hbar}{2}\Oo(t)\cot\phi(t),
  \end{gather}
\end{subequations}
and the relative instantaneous eigenstates are
\begin{subequations}
  \label{eq:eigenstates}
  \begin{gather}
    \ket{a_0(t)} = \cos\theta(t)\ket{g} - \sin\theta(t)\ket{r}, \label{eq:a0}\\
    \begin{aligned}
      \ket{a_-(t)} ={} &\sin\theta(t)\cos\phi(t)\ket{g}-\sin\phi(t)\ket{e}+{}\\
      &+ \cos\theta(t)\cos\phi(t)\ket{r},
    \end{aligned}\\
    \begin{aligned}
      \ket{a_+(t)} ={} &\sin\theta(t)\sin\phi(t)\ket{g}+\cos\phi(t)\ket{e}+{}\\
      &+ \cos\theta(t)\sin\phi(t)\ket{r}.
    \end{aligned}
  \end{gather}
\end{subequations}
As the three-level key feature, the $\ket{a_0}$ eigenstate with eigenvalue zero
is a \textit{dark state}~\cite{ArimondoLaNC11976nonabsorbing} with zero
projection on state $\ket{e}$.

\subsubsection{STIRAP}
The STIRAP protocol allows for an efficient population transfer from state
$\ket{g}$ to state $\ket{r}$ by adiabatically following the dark state
$\ket{a_0}$. Since the state $\ket{a_0}$, eq.~\eqref{eq:a0}, does not have any
component along the excited state $\ket{e}$, the population losses at rate
$\gamma$ from that state have very little impact on the evolution of the system
and thus on the fidelity $\F$ of the process.
\begin{figure}[!ht]
  \centering
  \includegraphics[width=0.93\linewidth]{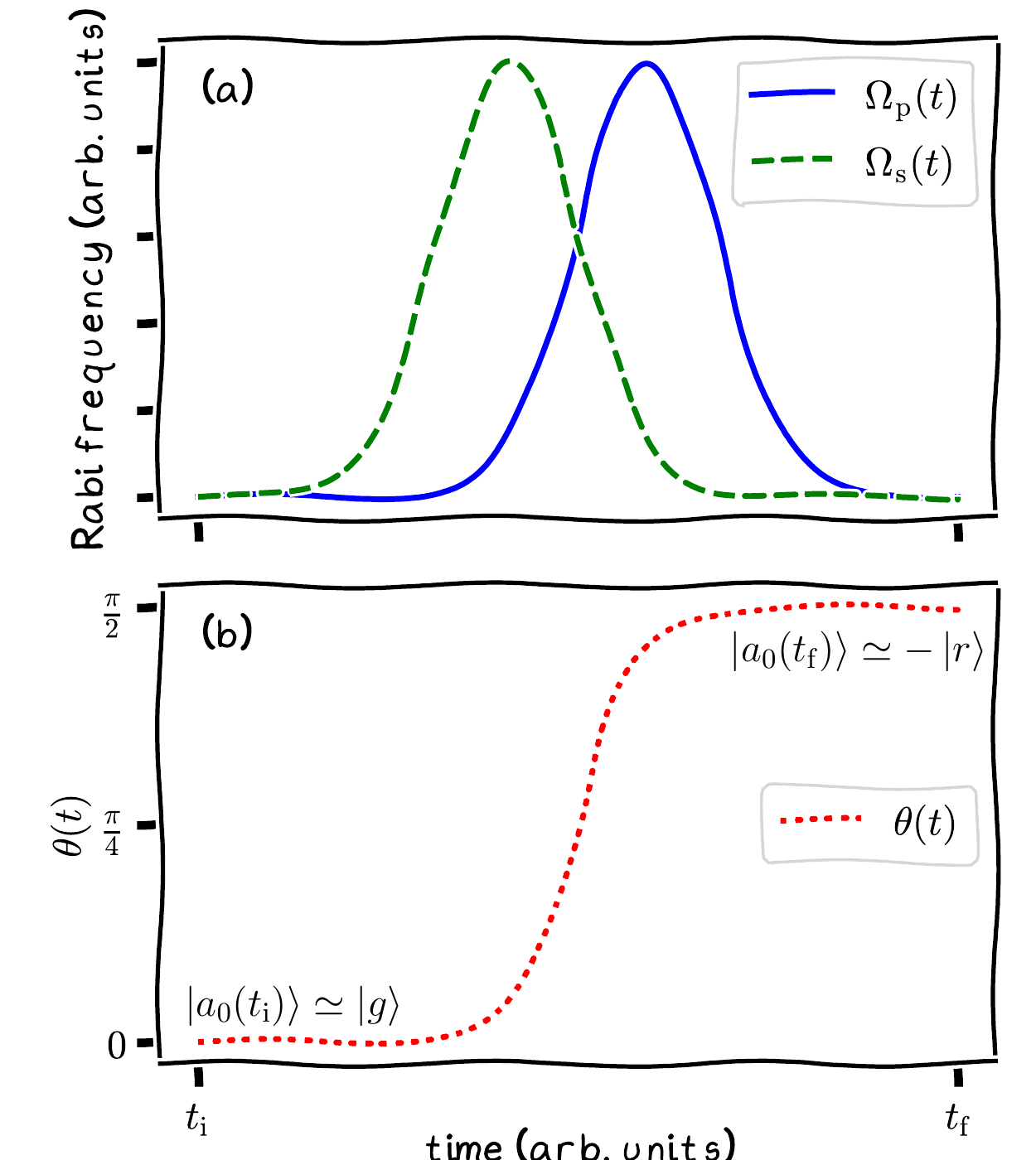}
  \caption{\label{fig:samplepulses}Counter-intuitive sequence of pulses
    peculiar to the STIRAP protocol. (a) Pump and Stokes gaussian pulses
    of the type $\Omega_\mathrm{p/s}(t) \propto \e^{-\brr{(t\mp\tau)/T}^2}$,
    and (b) time dependence of $\theta(t)$ given by eq.~\eqref{eq:tantheta}
    with the Rabi frequencies of (a).}
\end{figure}

In order to achieve population transfer from state $\ket{g}$ to state $\ket{r}$
by following $\ket{a_0(t)}$ we need
\begin{subequations}
  \begin{align}
    \label{eq:a_0initialandfinalneeds}
    \ket{a_0(\ti)} &\propto \ket{g}, \\
    \ket{a_0(\tf)} &\propto \ket{r},
  \end{align}
\end{subequations}
where $\ti$ and $\tf$ are the initial and final time, respectively. This is
obtained if the Stokes and pump pulses satisfy
\begin{subequations}
  \label{eq:limtantheta}
  \begin{gather}
    \lim_{t\to \ti}\tan\theta(t) = \lim_{t\to \ti}\frac{\Op(t)}{\Os(t)} = 0, \\
    \lim_{t\to \tf}\cot\theta(t) = \lim_{t\to \tf}\frac{\Os(t)}{\Op(t)} = 0,
  \end{gather}
\end{subequations}
and, equivalently
$\lim_{t\to \ti}\theta(t)=0,\ \lim_{t\to \tf}\theta(t)=\pm \pi/2$.

In order for the evolution to be adiabatic, the pulses must also satisfy the
condition~\cite{BergmannRMP1998coherent,FleischhauerPRA1996propagation}
\begin{equation}
  \label{eq:STIRAPlocaladiabaticondition}
  \vert\dot\theta(t)\vert \ll \frac{1}{2}\abs{\Dp \pm\sqrt{\Dp^2 + \Oo(t)^2}},
\end{equation}
which is obtained by applying eq.~\eqref{eq:adiabaticondition} to the
eigensystem given in eqs.~\eqref{eq:eigenvalues}
and~\eqref{eq:eigenstates}. Eq.~\eqref{eq:STIRAPlocaladiabaticondition} is a
\textit{local} adiabaticity condition and must be valid for every time $t$.

Eqs.~\eqref{eq:limtantheta} and~\eqref{eq:STIRAPlocaladiabaticondition}
mathematically formalize the concept of counter-intuitive pulse sequence
peculiar to the STIRAP protocol: the Stokes pulse (which couples the initially
empty states $\ket{e}$ and $\ket{r}$) is applied first, then it gets slowly
turned off while the pump pulse is turned on, having an overlap with the Stokes
pulse. Being and adiabatic protocol, STIRAP is very robust against noise in the
control fields~\cite{VitanovRMP2017stimulated}.

Typically used pulses are of the form
\begin{equation}
  \label{eq:typicalpulses}
  \Op(t) = \Omega_\mathrm{max}f\br{\frac{t-\tau}{T}},\quad
  \Os(t) = \alpha\Omega_\mathrm{max}f\br{\frac{t+\tau}{T}},
\end{equation}
where $f(t)$ is a pulse envelope having unit maximum value,
$\Omega_\mathrm{max}$ is the peak Rabi frequency, $2\tau$ is the delay between
the pulses, $T$ is the pulse width, and $\alpha$ is a scaling parameter usually
equal to $1$. The counter-intuitive sequence condition imposes $\tau>0$.

By assuming $\Dp\ll\Op(t),\Os(t)$, a global adiabaticity condition is derived
by time averaging eq.~\eqref{eq:STIRAPlocaladiabaticondition} over the
characteristic time $\tau$ of the $\Op(t)$ and $\Os(t)$ overlap. For the pulses
of eq.~\eqref{eq:typicalpulses} using eqs.~\eqref{eq:limtantheta} the global
adiabaticity condition becomes\footnote{Often the condition
  used\cite{BergmannRMP1998coherent} is $\Omega_\mathrm{max}\tau\geq 10$.}
$\Omega_\mathrm{max}\tau\gg1$.

An example of STIRAP pulses is plotted in Fig.~\ref{fig:samplepulses}. For a
list of various pulse shapes used in literature and the relative superadiabatic
solution we refer to~\cite{GiannelliPRA2014threelevel,PetiziolSR2020optimized}.

Finding new STIRAP-like protocols is important in solid-state systems where
often one must operate with reduced control
resources~\cite{DiStefanoPRB2015population,DiStefanoPRA2016coherent} since they
provide new tools for coherently
probing~\cite{FalciFP2017advances,FalciSR2019ultrastrong} or for
processing~\cite{RidolfoEPJST2021probing} in quantum architectures.

\section{Optimal Control\label{sec:optimal-control}}
In this section we introduce the \textit{Optimal Control problem} and one way to
approach its solution numerically. For a more rigorous and general treatment we
refer the reader to the classical
books~\cite{dAlessandro2021introduction,Liberzon2012calculus} and the
introductory
reviews~\cite{BoscainPRXQ2021introduction,RemboldAQS2020introduction,WilhelmAQ2020introduction}.

\subsection{Formulation of the Optimal Control problem}
Consider a system described by the (non-linear) set of differential equations
\begin{equation}
  \label{eq:OCTproblem}
  \begin{gathered}
    \dot\rho(t) = f\br{\rho(t),\vect{u}(t),t}, \quad t\in[0,T], \\
    \vect{u}(t) = \br{u_1(t), u_2(t), \dots, u_M(t)},
  \end{gathered}
\end{equation}
where $\rho(t)$ represents the state of the system, $f$ is a \textit{smooth
  function} which describes the dynamics of the state $\rho(t)$ and depends on
the $M$ control functions $\vect{u}(t)$. The objective of \textit{optimal
  control} is to find some control functions $\vect{u}(t)$ such that the
dynamics of the system is as close as possible to the desired dynamics
(examples of what this means will be explicitly given later). This is done by
introducing a cost functional
\begin{equation}
  \label{eq:OCTcostfunction}
  \mathcal{J}(\rho(t), \vect{u}(t), T)
\end{equation}
whose minimization corresponds to the desired dynamics.

\subsubsection{Quantum Optimal Control}
\textit{Quantum Control} consists in the control of the evolution of a quantum
system. We can formulate the quantum control problem as
\begin{equation}
  \label{eq:QCproblem}
  \dot{\rho}(t) = \Lio(t,\vect{u}(t))\rho(t),
\end{equation}
\ie, by identifying in eq.~\eqref{eq:OCTproblem} $\rho(t)$ with the density
matrix, $f$ with a super-operator that acts on the space of density matrices
and that describes the time evolution of the system, and $\vect{u}(t)$ with some
external controls. The time evolution of the system can be
expressed as
\begin{equation}
  \label{eq:quantumtimeevolution}
  \rho(t) = \U\rho(0),
\end{equation}
where with $\U$ we indicate the time-evolution superoperator (or
\textit{quantum map}) which does not need to be unitary since it can describe
both the coherent and incoherent dynamics~\cite{KochJPCM2016controlling}.

The \textit{Quantum Optimal Control problem} consists in determining the control
amplitudes $\vect{u}(t)$ that will perform the quantum operation of interest,
\eg drive the system from the given initial state $\rho(0)$ at time $t=0$ to the
target state $\rho_\mathrm{targ}$ at time $t=T$ (this process is called
\textit{state transfer}), or that implements a transformation $R_\mathrm{targ}$
in the time interval $[0,T]$ (\textit{gate synthesis})\footnote{Other
  possibilities are maximizing the entanglement
  generation~\cite{WattsPRA2015optimizing,GoerzPRA2015optimizing,MullerPRA2011optimizing},
  or state distinguishability~\cite{BasilewitschPRR2020optimally}, to name a
  few.}.

To quantify how close the evolution given by $\U$ is to the target evolution we
define a cost functional $\mathcal{J}$ (as in eq.~\eqref{eq:OCTcostfunction})
that we seek to minimize, with respect to the controls
$\vect{u}(t)$. $\mathcal{J}(\rho(t), \vect{u}(t), T)$ being minimal should
correspond to the ideal process we want to perform. Commonly used functionals
for quantum processes are of the form
\begin{equation}
  \label{eq:OCTfunctional}
  \mathcal{J}(\rho(t), \vect{u}(t), T) = 1 - \F
\end{equation}
where $\F$ is the transfer
fidelity~\cite{KhanejaJMR2005optimal,RemboldAQS2020introduction,WilhelmAQ2020introduction}
\begin{equation}
  \label{eq:OCTfidelity}
  \F = \Tr\brrr{\rho_\mathrm{targ}^\dag\rho(T)},
\end{equation}
for the case of state transfer, and the gate
fidelity~\cite{PalaoPRA2003optimal,MontangeroPRL2007robust,SaidPRA2009robust,GoerzPRA2015optimizing,RemboldAQS2020introduction,WilhelmAQ2020introduction,GoerzNJP2014optimal}
\begin{equation}
  \label{eq:OCTgatesynthesisfid}
  \F = d^{-1}\abs{\Tr\brrr{R_\mathrm{targ}^\dag R(T,\vect{u}(t))}},
\end{equation}
where $d$ is the dimension of the Hilbert space of the system, for the case of
gate synthesis.

Notice that the final time $T$ can be fixed, or can be included in the
functional in order to minimize also the duration of a quantum process.

Constraints on the controls can also be included in the definition of the cost
functional
$\mathcal{J}$~\cite{BoscainPRXQ2021introduction,RemboldAQS2020introduction,WilhelmAQ2020introduction,PalaoPRA2008protecting}
but we do not consider them here.

\subsubsection{Hamiltonian Control}
A typical situation encountered in quantum control is when each control
amplitude $u_j(t)$ in $\vect{u}(t)$ corresponds to an external tunable
parameter which can be described by Hermitian operator in the Hamiltonian. For
the purpose of this tutorial we assume that the dynamics of the system can be
described by the Lindblad master
equation~\cite{GoriniJMP1976completely,LindbladCMP1976generators}
\begin{equation}
  \label{eq:OCTmastereq}
  \dot{\rho}(t) = \Lio(t)\rho(t) = -\I [H(t),\rho(t)] + \Lio_\mathrm{dis}\rho(t),
\end{equation}
where the Hamiltonian $H(t)$ describes the coherent dynamics and
$\Lio_\mathrm{dis}$ the incoherent dynamics. The Hamiltonian can be written as
\begin{equation}
  \label{eq:OCTHamiltonian}
  H(t) = H_0 + \sum_{k=1}^M u_k(t)H_k,
\end{equation}
where $H_0$ is the free evolution Hamiltonian (often called \textit{drift}
Hamiltonian), $H_k$ for $k=1,\dots,M$ are the available control Hamiltonians
corresponding to operations on the system we can control, and $u_k(t)$ are the
time-varying amplitude functions for their relative control.  The solution of
eq.~\eqref{eq:OCTmastereq} can be written as
\begin{subequations}
  \label{eq:OCTequation}
  \begin{gather}
    \rho(t) = \U[\rho(0)], \\
    \U = \mathcal{T}\exp{\int_0^t\Lio(t')\de t'} \\
    \vect{u}(t) = \br{u_1(t), u_2(t), \dots, u_M(t)},
  \end{gather}
\end{subequations}
and $\mathcal{T}$ is the time-ordering operator.

To summarize, we now want to find a set of controls $\vect{u}(t)$ such that the
evolution of the system given by eqs.~\eqref{eq:OCTequation} is the target
evolution. To quantify how close the system
evolution is to the target one, we need to minimize the cost functional in
eq.~\eqref{eq:OCTfunctional} with eqs.~\eqref{eq:OCTfidelity}
or~\eqref{eq:OCTgatesynthesisfid}.

Once the problem has been defined, a method for minimizing the cost functional
$\mathcal{J}$ is required. Various strategies exist to solve this problem, both
analytical methods based on calculus of variations and the Pontryagins minimum
principle~\cite{BoscainPRXQ2021introduction}, and numerical methods. In this
tutorial we focus exclusively on numerical methods.

\subsection{Numerical solution of the Quantum Optimal Control
  problem\label{sec:numer-solut-quant}}
The numerical solution of the Quantum Optimal Control problem requires the
mapping of the cost function $\mathcal{J}$, eq.~\eqref{eq:OCTcostfunction}, to
a multivariate real function $\bar{\mathcal{J}}$, and then numerically minimize
$\bar{\mathcal{J}}$.

The mapping is done by parametrizing each control amplitude $u_k(t)$ with $N_k$
real numbers, \ie with a vector $\ak\in\mathbb{R}^N_k$. This is done by
approximating\footnote{The parametrization reduces the space of functions in
  which we look for a solution, so it is important to have a good
  parametrization if we want to find a quasi-optimal solution.} each $u_k(t)$
as an expansion on a finite set of functions as in
\begin{equation}
  \label{eq:OCTexpasion}
  u_k(t) = \sum_{j=1}^{N_k'} c_{kj}\chi_{kj}(t,\vect{d}_{kj}),
\end{equation}
where $\chi_{kj}(t,\vect{d}_{kj})$ are time-dependent functions which depend on
the parameters $\vect{d}_{kj}$, and
\begin{equation*}
  \ak = \{\Re\br{c_{kj}}, \Im\br{c_{kj}}, \Re\br{\vect{d}_{kj}},
  \Im\br{\vect{d}_{kj}}\}_{j=1,\dots,N_k'}
\end{equation*}
are the $N_k$ parameters representing the function $u_k(t)$. Notice that we can
choose a different set of function $\chi_{kj}$ for each control $u_k(t)$.
\begin{figure}[!ht]
  \centering
  \includegraphics[width=0.93\linewidth]{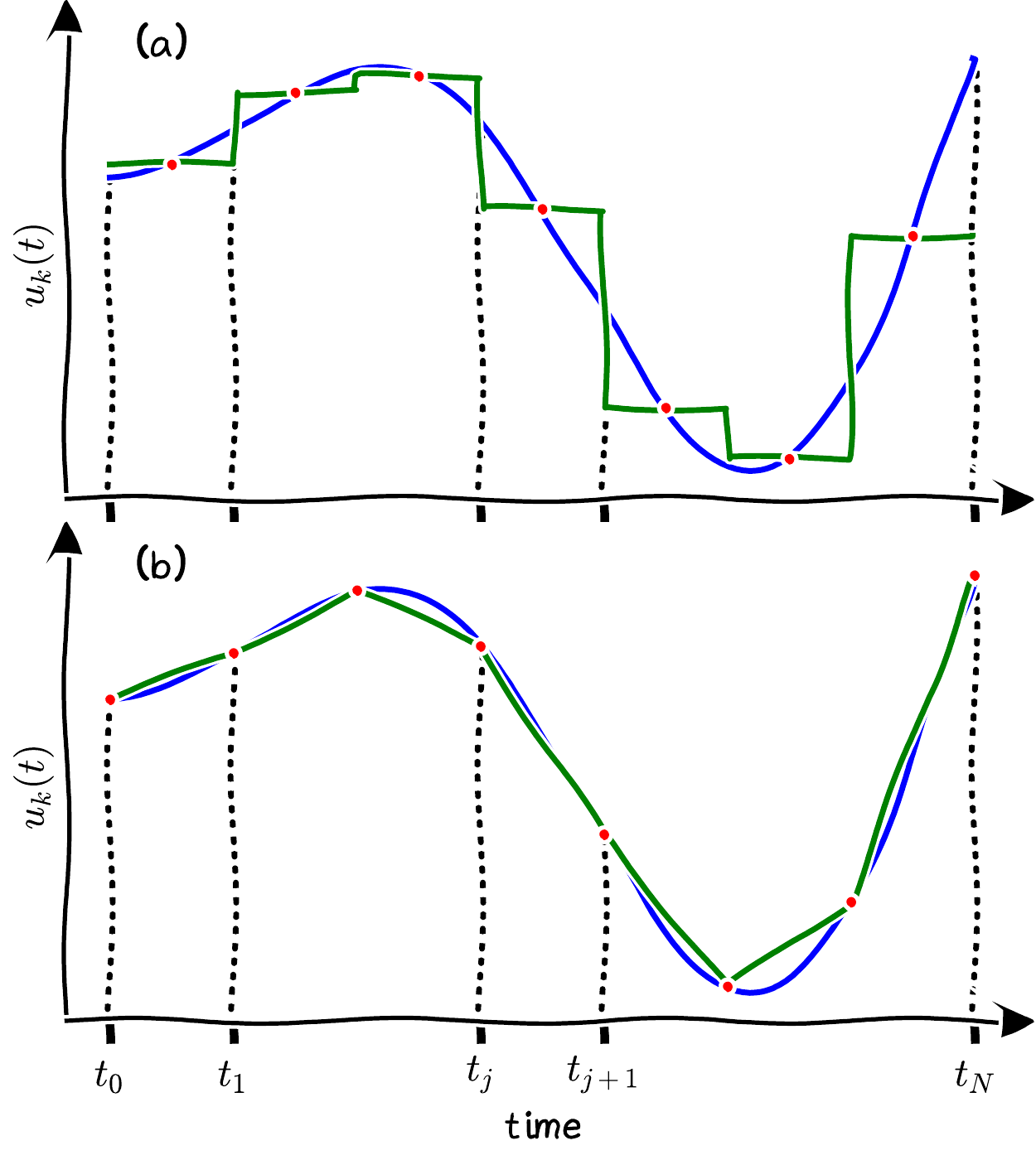}
  \caption{\label{fig:intervalsplitting}Artistic view of piecewise (a) constant
    and (b) linear functions that can approximate the real control amplitude
    $u_k(t)$.}
\end{figure}

A useful and simple set of functions which can be used to expand the control
amplitudes are piecewise functions: we split the time interval $[0,T]$ in $N'$
smaller intervals $I_j$ of length $\Delta t = T/N'$ as in the following:
\begin{equation}
  \label{eq:timeintervals}
  I_j = [t_{j-1},t_j), \quad j=1,\dots,N',
\end{equation}
such that $t_0=0$, $t_{N'}=T$, and $t_j = t_{j-1}+\Delta t$. With this time
discretization we can define
\begin{equation}
  \label{eq:OCTchikj}
  \chi_{kj}(t,\vect{d}_{kj}) =
  \begin{cases}
    f_{kj}(t,\vect{d}_{kj}) & t\in I_j \\
    0 & \text{otherwise}
  \end{cases}
\end{equation}
where $f_{kj}$ are some (possibly) time-dependent functions\footnote{Here we
  assume that we use the same number of time steps for each control amplitude
  $u_k(t)$, thus $N'_k = N'$.}. The functions $\chi_{kj}(t,\vect{d}_{kj})$ are
different from zero only in the interval $I_j$, such that the function $u_k$ is
equal to the function $f_{kj}(t,\vect{d}_{kj})$ in the interval $I_j$.

An important set of functions often used are \textit{step functions} (or
\textit{piecewise constant functions}): In some problems (such as state
transfer and gate synthesis as described above) they decrease the computational
cost of calculating the gradient~\cite{KhanejaJMR2005optimal} and thus speed up
the numerical minimization using gradient-based algorithms. They are obtained
by choosing $f_{kj}(t,\vect{d}_{kj}) = 1$\footnote{With this choice each
  control amplitude is expanded on the same set of function and thus is
  represented by the same number of parameters $N_k=N$.} in
eq.~\eqref{eq:OCTchikj}, then $u_k(t)$ can be easily written as
\begin{equation}
  \label{eq:OCTstepfunction}
  u_k(t) =
  \begin{cases}
    c_{kj} & t\in I_j \\
    0 & \text{otherwise}.
  \end{cases}
\end{equation}
The $N=2N'$ real parameters representing the function $u_k(t)$ can be chosen to
be
\begin{equation*}
  \ak = (\Re\br{c_{k1}}, \dots, \Re\br{c_{kN'}}, \Im\br{c_{k1}}, \dots,
  \Im\br{c_{kN'}}).
\end{equation*}
If we assume the function $u_k(t)$ to be real, then $N=N'$ and
$\ak = (c_{k1}, c_{k2}, \dots, c_{kN'})$. Notice that
eq.~\eqref{eq:OCTstepfunction} is the nearest-neighbor constant interpolation
of the points $\ak = (c_{k1}, c_{k2}, \dots, c_{kN'})$, see
Fig.~\ref{fig:intervalsplitting}(a).

If $f_{kj}$ are linear functions and we impose $u_k(t)$ to be continuous, then
\begin{equation}
  \label{eq:OCTlinearfunction}
  f_{kj}(t,\vect{d}_{kj}) = \frac{c_{k(j-1)}}{c_{kj}} +
  \frac{c_{kj}- c_{k(j-1)}}{c_{kj} \Delta t}(t-t_{j-1}),
\end{equation}
with $\vect{d}_{kj}=\{c_{k(j-1)},c_{kj}\}$. In this case the control
function $u_k(t)$ can be easily written as
\begin{equation}
  \label{eq:OCTlinearexpansion}
  u_k(t) =
  \begin{cases}
    c_{k(j-1)} + \frac{c_{kj}- c_{k(j-1)}}{\Delta t} (t-t_{j-1}) & t\in I_j \\
    0 & \text{otherwise}.
  \end{cases}
\end{equation}
If we assume that the function $u_k(t)$ is real\footnote{Every discussion can
  be easily extended to complex functions by considering each complex parameter
  as two real parameters.}, then it can be parametrized by $N=N'+1$ real
parameters $\ak = (c_{k0}, c_{k1}, \dots, c_{kN'})$.  Notice that
eq.~\eqref{eq:OCTlinearexpansion} is the linear interpolation of the points
$\ak = (c_{k0}, c_{k1}, \dots, c_{kN'})$, see
Fig.~\ref{fig:intervalsplitting}(b).

\subsubsection{Final minimization}
Once we have chosen a parametrization of the functions
$\vect{u}(t)$\footnote{In general we can choose a different parametrization for
  each function $u_j(t)$ in $\vect{u}(t)$, also with a different number of
  parameters $N_j$ for each function. For the sake of presentation we report
  the case in which the number of parameters $N_j=N$ is the same for each
  function.} we collect all the parameters $\ak \in \mathbb{R}^N$ of all the
functions $u_k(t)$ for $k=1,\dots,M$ into a single vector
$\vect{\alpha} \in \mathbb{R}^{N\times M}$ so that we can write the cost
functional as
\begin{equation}
  \label{eq:OCTcostfunctional}
  \mathcal{J}(\rho(t), \vect{u}(t), T) = \bar{\mathcal{J}}(\rho(t),
  \vect{\alpha}, T)
\end{equation}
where in $\bar{\mathcal{J}}$ the evolution of $\rho(t)$ is computed with the
control amplitudes parametrized by $\vect{\alpha}$. Now it is possible to
minimize $\bar{\mathcal{J}}(\rho(t), \vect{\alpha}, T)$ with respect to the
$N\times M$ real parameters using any of the numerical methods developed to
minimize multivariate real functions.

An issue that is often encountered in the minimization process consists in the
algorithm being stuck in a local minimum: usually the numerical algorithms will
find the local minimum which is the closest to the starting point (called
\textit{initial guess}). While increasing the number of parameters can
potentially solve this problem~\cite{RabitzS2004quantum}, several methods have
also been developed in order to address this issue, see for
example~\cite{GoerzEQT2015hybrid,GoerznQI2017charting,BasilewitschAQ2021engineering}.
A simple approach is to try different initial guesses and choose the minimization
which gives the minimum value of $\bar{\mathcal{J}}$.

Several methods have been developed in order to speed up the minimization of the
cost function $\bar{\mathcal{J}}$. They use properties of the dynamics of the
systems in order to decrease the computational cost of calculating its gradient,
or choose a suitable set of functions $\chi_{kj}$ in order to speed up the
convergence of the minimization algorithm, or to reduce the dimensionality of
the optimization problem~\cite{LucarelliPRA2018quantum}. Here we list some of
the most common algorithms, while we refer the reader to the original paper or
the recent reviews~\cite{RemboldAQS2020introduction,WilhelmAQ2020introduction}
for a deeper explanation: \textit{GRAPE}~\cite{KhanejaJMR2005optimal},
\textit{Krotov}~\cite{KrotovAiNDaCARfR1993global,KonnovARC1999global,SklarzPRA2002loading,ReichJCP2012monotonically},
\textit{GOAT}~\cite{MachnesPRL2018tunable},
\textit{CRAB}~\cite{DoriaPRL2011optimal,CanevaPRA2011chopped,MullerAQ2021one},
\textit{dCRAB}~\cite{RachPRA2015dressing}.

\subsection{Three-level population transfer}
We now formulate the three-level population transfer process introduced in
section~\ref{sec:three-level-popul} as an Optimal Control problem and solve it
numerically.

Following eq.~\eqref{eq:OCTHamiltonian} we identify in
eq.~\eqref{eq:threelevelHamiltonian} the \textit{drift Hamiltonian} ($\hbar=1$)
as
\begin{equation}
  \label{eq:3LPT-drifthamiltonian}
  H_0 = \Dp\ketbra{e}{e} + \D\ketbra{r}{r}
\end{equation}
and the \textit{control Hamiltonians} as
\begin{align}
  H_1 ={} &\frac{\ketbra{g}{e} + \ketbra{e}{g}}{2} \\
  H_2 ={} &\frac{\ketbra{e}{r} + \ketbra{r}{e}}{2}
\end{align}
with $u_1(t) = \Op(t)$ and $u_2(t) = \Os(t)$. We assume the controls
$\Op(t)$ and $\Os(t)$ to be real constant piecewise functions and parametrize
each of them with $N$ parameters corresponding to the value they assume on each
time interval, see sec.~\ref{sec:numer-solut-quant} and in particular
eqs.~\eqref{eq:timeintervals} and~\eqref{eq:OCTstepfunction}. We collect the
$2N$ parameters in the vector $\vect{\alpha} \in \mathbb{R}^{2N}$.

Our goal is to find the control amplitudes $\Op(t)$ and $\Os(t)$ that maximize
the population on the state $\ket{\psi_\mathrm{targ}} = \ket{r}$ at time $t=T$
starting from the state $\ket{\psi(0)}=\ket{g}$ at time $t=0$. Thus we define
the cost function $\mathcal{J}(\rho(t), \vect{u}(t), T)$ as in
eqs.~\eqref{eq:OCTfunctional} and~\eqref{eq:OCTfidelity}, \ie:
\begin{equation}
  \label{eq:2LPT-costfunction}
  \bar{\mathcal{J}}(\rho(t), \vect{\alpha}, T) =
  \mathcal{J}(\rho(t),\vect{u}(t), T) =
  1 - \Tr\brrr{\rho_\mathrm{targ}^\dag\rho(T)},
\end{equation}
where
$\rho_\mathrm{targ} = \ketbra{\psi_\mathrm{targ}}{\psi_\mathrm{targ}} =
\ketbra{r}{r}$ with $\rho(T)$ obtained from the evolution following the master
equation~\eqref{eq:mastereq} (equivalently eq.~\eqref{eq:OCTmastereq}) with the
initial density matrix being $\rho(0) = \ketbra{g}{g}$.

\subsubsection{Results}
We minimize $\bar{\mathcal{J}}(\rho(t), \vect{\alpha}, T)$ with respect to
$\alpha$ numerically, with $N=30$. Since the problem is easy (it consists in solving
numerically a system of $16$ coupled linear differential equations, which we do
numerically with \textit{QuTiP}~\cite{JohanssonCPC2013qutip}) we do not
use any advanced method. We have tried the \textit{Powell
  method}~\cite{PowellTCJ1964efficient}, the \textit{Nelder-Mead
  algorithm}~\cite{NelderTCJ1965simplex}, and the \textit{limited memory BFGS
  bounded} (\textit{L-BFGS-B}) method\footnote{With the gradient computed
  numerically.}~\cite{ByrdSJSC1995limited} as implemented by
\textit{SciPy}~\cite{VirtanenNM2020scipy}. We present the results obtained with
\textit{L-BFGS-B} since we have found that it is the fastest (with
\textit{Nelder-Mead} being the slowest).

The maximum efficiency of the protocol depends on the constraints given by the
time of the transfer $T$, the decay rate $\gamma$, and the maximum allowed Rabi
frequency $\Om$. However, since there is a freedom on the choice of the unit of
time (or equivalently the unit of frequency), the system is invariant under a
transformation that keeps $T\gamma$ and $T\Om$ constant, \ie, if
\begin{equation}
  \label{eq:invarianttransformation}
  \begin{cases}
    T' = \alpha T, \\
    \gamma' = \gamma/\alpha, \\
    \Om' = \Om/\alpha,
  \end{cases}
\end{equation}
then the system with parameters $(T,\gamma,\Om)$ is mathematically equivalent
to the system with $(T',\gamma',\Om')$. In particular the fidelities
$\F(T,\gamma,\Om) = \F(T',\gamma',\Om') = \F(T\gamma,T\Om)$ are the same. Thus
in the following we will report the results as a function of $T\gamma$ and
$T\Om$.

\begin{figure}[!ht]
  \centering
  \input{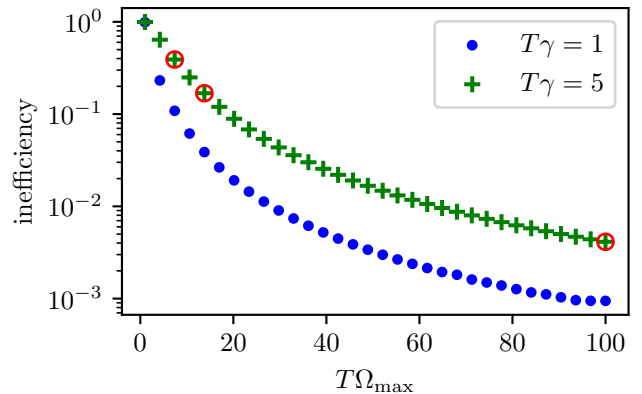}
  \caption{Inefficiency $1 - \F$, eq.~\eqref{eq:OCTfidelity} of the optimized
    protocol, as a function of $T\Om$, for various values of $T\gamma$. For
    each point we have optimized with $4$ different initial guesses and chosen
    the best one. The $3$ red empty circles refer to the parameters used in
    Fig.~\ref{fig:example_OCT_gamma5}.}
  \label{fig:ineff_vs_Omax}
\end{figure}

Fig.~\ref{fig:ineff_vs_Omax} reports the inefficiency of the optimized protocol
with respect to $T\Om$ for various values of the decay rate $T\gamma$.
\begin{figure*}[!ht]
  \centering
  \input{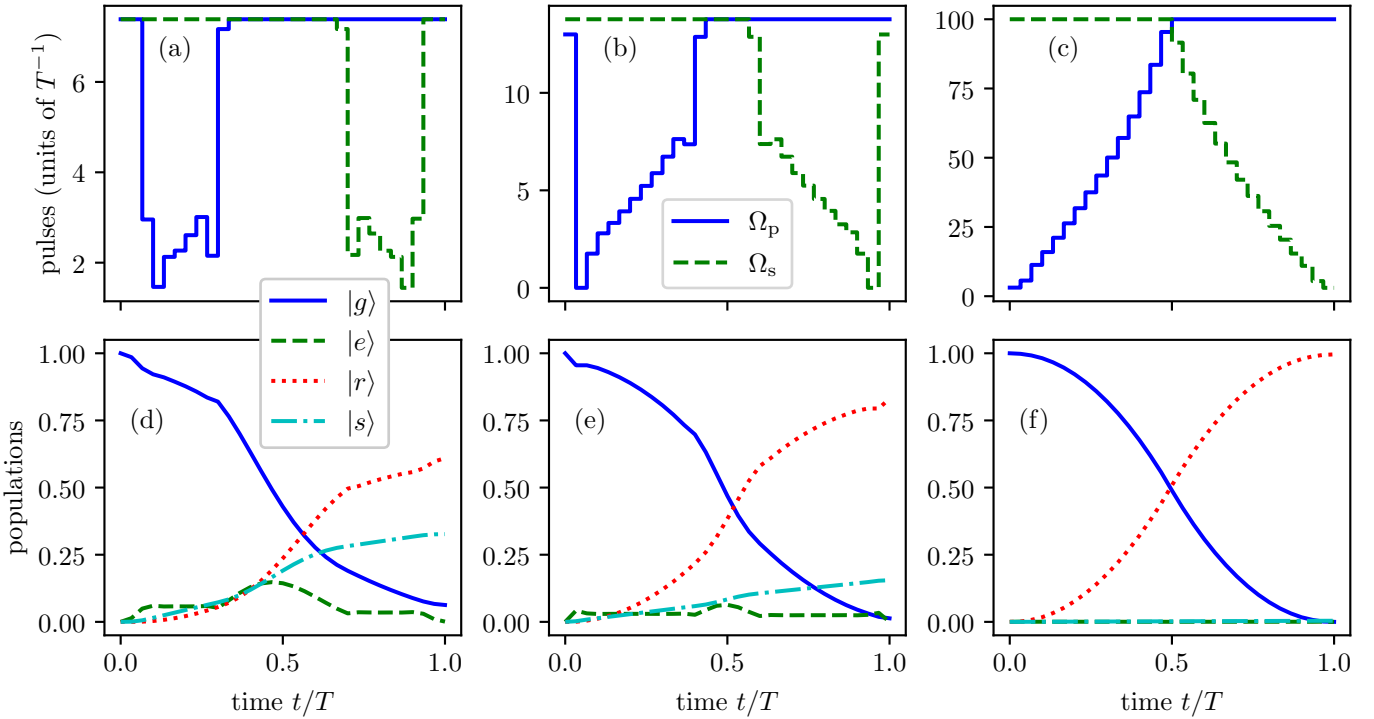}
  \caption{Example of optimized pulses and evolution for $T\gamma=5$. The top
    panels (a-c) report the optimized pulses $\Op(t)$ and $\Os(t)$, the lower
    panels (d-f) report the population of the states of the system when driven
    by the pulses in (a-c), respectively. For (a) and (d) $T\Om = 7.4$, for (b)
    and (e) $T\Om = 13.8$, and for (c) and (f) $T\Om = 100$. We have also
    chosen $\Op(t)$ and $\Os(r)$ to be real and $N=30$.}
  \label{fig:example_OCT_gamma5}
\end{figure*}
The red circles on the $T\gamma=5$ line refer to the points represented in
Fig.~\ref{fig:example_OCT_gamma5}. For all values of $T\Om$ the optimized
pulses recall the typical counter-intuitive pulse sequence of STIRAP: for
values of $T\Om\lesssim 40$ the optimized pulses present an initial and final
maximum interleaved by the counter-intuitive sequence. The area of this initial
and final \textit{short bumps} decreases with increasing $T\Om$. For
$T\Om\gtrsim 40$ the pulses are exactly counter-intuitive and they tend to
maximize the area at their disposal and their overlap still meeting the
condition of being counter-intuitive and the constraint
$\Omega_\mathrm{p,s}\leq\Om$. They do so by being symmetric with respect to the
central time $t=T/2$ and linearly increasing ($\Op$) or decreasing ($\Os$).

In circuit quantum electrodynamic systems, microwave pulses with these shapes
can be realized experimentally by using programmable arbitrary waveform
generators and mixers; this has been used already for example to demonstrate
STIRAP~\cite{KumarNC2016stimulated} and superadiabatic
STIRAP~\cite{VepsalainenSA2019superadiabatic,VepsalainenAQT2020simulating}.

\section{Reinforcement learning\label{sec:reinf-learn}}
Due to their wide range of applicability and their recent overwelming success
when used in combination with Deep Neural
Networks~\cite{HornikNN1989multilayer}, Reinforcement Learning (RL) techniques
have gathered significant interest at both the academic and industrial level
across a multitude of disciplines. Deep Reinforcement Learning (DRL) has already
provided several outstanding results such as solving complex continuous control
tasks~\cite{LillicrapACS2019continuous}, video game
play~\cite{MnihAC2013playing,MnihN2015humanlevel} and mastering the game of
Go~\cite{SilverN2017mastering} to highlight only a small handful. More recently,
DRL has emerged as a useful tool for quantum technologies and in particular has
provided a viable alternative strategy for solving Quantum Optimal Control
problems. RL has thus far been applied to quantum systems in the context of
state preparation~\cite{sivak2021model, haug2020classifying,PorottiAQ2021deep},
circuit architecture design~\cite{kuo2021quantum}, quantum
control~\cite{NiunQI2019universal,an2021quantum,BorahPRL2021measurementbased,FallaniAQ2021learning},
state transfer~\cite{PaparellePLA2020digitally,
  PorottiCP2019coherent,BrownNJP2021reinforcement}, quantum noise detection and
correction~\cite{MavadiaNC2017prediction,FoselPRX2018reinforcement,NiunQI2019universal},
quantum compiling~\cite{MoroCP2021quantum} and entropy production in
non-equilibrium quantum thermodynamics~\cite{SgroiPRL2021reinforcement}.

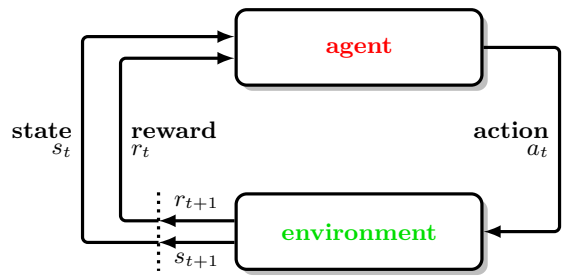
\begin{figure}[!ht]
  \centering
  \begin{tikzpicture}[font=\small\bfseries,node distance = 1cm, auto, very thick]
  \tikzstyle{block} = [rectangle, draw, text width=3cm, text centered, drop
  shadow, fill=white, centered, rounded corners, minimum height=1cm]
  \tikzstyle{line} = [draw, -{latex}, rounded corners=0.7mm]

  \node[block] (agent) {\color{red}agent};

  \node[block, below=1.4cm of agent] (environment)
  {\color[rgb]{0,0.87,0}environment};

  \path[line] (agent.0) --++ (1cm,0) |- node[near start, left, text
  width=1.2cm, align=right]{action\\[-0.17cm] $a_t$} (environment.0);

  \coordinate[left=10mm of environment] (P); \draw[dotted]
  (P|-environment.north) -- (P|-environment.south);

  \path[line] (environment.185) -- (P|-environment.185) node[midway, below]
  {$s_{t+1}$};

  \path[line] (environment.175) -- (P|-environment.175) node[midway, above]
  {$r_{t+1}$};

  \path[line] (P|-environment.185) --++ (-1cm,0) |- node [near start, text
  width=1cm, align=right] {state\\[-0.17cm] $s_{t}$} (agent.175);

  \path[line] (P|-environment.175) --++ (-0.5cm,0) |- node [near start, text
  width=1.2cm, right] {reward\\[-0.17cm] $r_{t}$} (agent.185);
\end{tikzpicture}
  \caption{Agent-Envitonment interface.}
  \label{fig:AgentEnvironmentInterface}
\end{figure}

In a typical RL setting, an agent dynamically interacts with an environment
with the goal of performing a certain task. A set of discrete interactions
between the agent and the environment is usually assumed. During each of these
interactions, the agent observes the state of the environment and, based on
this observation, performs a certain action. The state of the environment for
the next interaction will depend on this action while the agent is provided
with a feedback (called \textit{reward}) based on how well it is
performing the assigned task. The reward can then be used
to update the agent's behaviour in order to improve its performance. A sketch
of this interaction is reported in Fig.~\ref{fig:AgentEnvironmentInterface}.

The process of sequential decision making that underpins RL is mathematically
formulated using so called Markov Decision Processes (MDPs). A full treatment of
MDPs does not fall in the remit of this tutorial, however in the following
section we will provide a condensed treatment which will be sufficient to then
introduce the specific RL algorithms of interest in a somewhat self contained
manner\footnote{For a full treatment of Markov Decision Processes in the context
  of Reinforcement Learning see the famed book of Sutton and
  Barto~\cite{Sutton2018reinforcement}. }.

\subsection{Markov Decision Processes}
Consider a general learning \textit{agent} that is able to repeatedly interact
with an \textit{environment} at discrete time-steps by first observing its
state, then taking actions that change this state. At the next time step the
new state and a reward are fed to the agent.

We define the state space $\mathcal{S}$, containing all conceivable states of
the environment, and an action space $\mathcal{A}(s)$, for all possible states
$s\in\mathcal{S}$. The reward is a scalar $r\in\mathcal{R}\subset\mathbb{R}$
and represents the performance of the agent. This interaction gives rise to a
\textit{trajectory}
\begin{equation}
  \label{eq:trajectory}
  S_0, A_0, R_1, S_1, A_1, R_2, \dots, S_t, A_t, R_t, \dots,
\end{equation}
where $S_0$ is the initial state of the environment, and $S_t$, $A_t$ and $R_t$
are the state, the action and the reward, respectively, at time step $t$. The
above decision process is said to be a Markov Decision Process (MDP) if the
state $S_t$ and reward $R_t$ at step $t$ depend only on the state $S_{t-1}$ and
action $A_{t-1}$ at step $t - 1$\footnote{This property is called \textit{Markov
    property.} Notice that this is not a restriction on the dynamics or the
  decision process, but a requirement on the representation of the state}. The
dynamics of the MDP is defined by the \textit{dynamic
  function}\footnote{Equation \ref{eq:dynamicfunction} refers to a
  \textit{finite MDP}, i.e to the case when $\mathcal{S}$, $\mathcal{A}(s)$ are
  finite sets. The formalism can be easily generalized to infinite state and
  action spaces.}~\cite{Sutton2018reinforcement}
\begin{equation}
  \label{eq:dynamicfunction}
  p(s', r | s, a) = \text{Pr}\{S_t=s', R_t=r | S_{t-1}=s, A_{t-1}=a\},
\end{equation}
which is the probability that at step $t$ the values of the state and reward
are $S_t = s'\in\mathcal{S}$ and $R_t=r\in\mathcal{R}$, given that at step
$t-1$ the values of the state and actions are $S_{t-1}=s\in\mathcal{S}$ and
$A_{t-1}=a\in\mathcal{A}(s)$\footnote{Notice that, since the MDPs considered in
  the review will model quantum system dynamics which in our case is
  deterministic, these probabilities can only takes values $0$ or $1$ and the
  reward will be simply a real function of $s$ and
  $a$.}~\cite{Sutton2018reinforcement}.

The rewards prescription must be representative of the desired goal (and thus be
sufficiently informative in that respect) whilst not including any information
about how the agent should go about achieving this goal (to avoid biasing
learning with already known strategies). In the case of quantum control as
defined in section~\ref{sec:numer-solut-quant}, the agent could be embodied by
the control mechanism employed in an experiment with the actions defined as the
choice of piecewise constant values of $u(t)$ at each discrete time-step. The
state of the MDP's environment would then be described by the state of the
quantum system at each time-step, for example via the density matrix. This
reformulation of the quantum control problem as an MDP is used
in~\cite{PorottiCP2019coherent,PaparellePLA2020digitally,BrownNJP2021reinforcement,SgroiPRL2021reinforcement}.

If the agent-environment interaction is interrupted after a terminal state
$S_N$ (enforced either by a maximum time or termination criteria) is reached,
then we will need to reset the environment state and start a new episode so
that the learning process may continue. In this case, the agent's task is said
to be \textit{episodic}. Otherwise, if the agent-environment interaction goes
on without limit, the task is referred to as \textit{continuing task}. Here we
exclusively consider episodic tasks.

The behaviour of a RL agent can be described with a conditional probability
distribution
\begin{equation}
  \pi(a|s)=\text{Pr}\{A_t=a|S_t=s\},
\end{equation}
usually referred as \textit{Policy} or \textit{Policy function}, that is, the
probability that the agent takes the action $A_t = a$ if the environment is
found in state $S_t=s$.

Consider now the trajectory of an episode for a MDP, as in
(\ref{eq:trajectory}). The \textit{return} for each time-step $t$ is defined as
\begin{equation}
  \label{eq:return}
  \begin{aligned}
    G_t &{}= \sum_{k=0}^{(N-t-1)} \Gamma^k R_{t+k+1} =\\
    &{}=R_{t+1} + \Gamma R_{t+2} + \Gamma^2 R_{t+3} \dots + \Gamma^{N-t-1} R_{N}.
  \end{aligned}
\end{equation}
which represents the ``discounted'' sum of future rewards. In equation
(\ref{eq:return}) the discount factor $\Gamma$ modulates the relative importance
of immediate versus future reward. For example, for $\Gamma=0$: $G_t = R_{t+1}$
which describes the situation where only immediate rewards are important. On the
other hand, for $\Gamma=1$: $G_t = \sum_{t} R_t$, so this return places equal
importance on immediate and future rewards.

The agent's performance can be evaluated from a certain state by looking at the
expected return. The expected return starting from the state $s$ and following
the policy $\pi$ (\ie, the remaining actions in the trajectory are selected
according to the policy $\pi$) is
\begin{equation}
  v_{\pi}(s)= \mathbb{E}_{\pi} [G_t|S_t=s].
\end{equation}
It is known as the \textit{state-value function} and satisfies the following
consistency condition (Bellman expectation Equation)
\begin{equation}
  v_{\pi}(s)= \sum_a \pi(a|s) \sum_{s', r}p(s', r|s, a)[r+\Gamma v_{\pi}(s')].
\end{equation}

The expected return starting from the state $s$, taking the action $a$ and
following the policy $\pi$
\begin{equation}
  q_{\pi}(s,a)= \mathbb{E}_{\pi} [G_t|S_t=s, A_t=a],
\end{equation}
is known as the \textit{action-value function}. A corresponding consistency
condition holds also for $q_{\pi}(s,a)$.

Any MDP admits one or more \textit{optimal policies} $\pi^*$ with optimal
value functions $v^*=\max_{\pi}v_{\pi}(s)$, $q^*=\max_{\pi}q_{\pi}(s,
a)$. Special consistency conditions, known as Bellman Optimality
Equations~\cite{Sutton2018reinforcement}, can be derived for the optimal value
functions

\begin{equation}\label{bellmanv}
  v^*(s)= \max_a\sum_{s', r}p(s', r|s, a)[r+\Gamma v^*(s')].
\end{equation}

\begin{equation}\label{bellmanq}
  q^*(s)= \sum_{s', r}p(s', r|s, a)[r+\Gamma \max_{a'}q^*(s', a')].
\end{equation}

In principle, one can find the exact solution of the Bellman Optimality
Equation \ref{bellmanv} to reconstruct the best policy with a one-step
search. However this is not usually possible for real world problems: even when
we have a complete model of the environment, it is usually not computationally
feasible to solve such equation.

Many successfull iterative solution methods have been developed based on
Equation \ref{bellmanq}. However, as we will show in the next section, one can
also approach the MDP from a different perspective without directly computing
any value function.

\begin{figure}[!ht]
 \centering
 \input{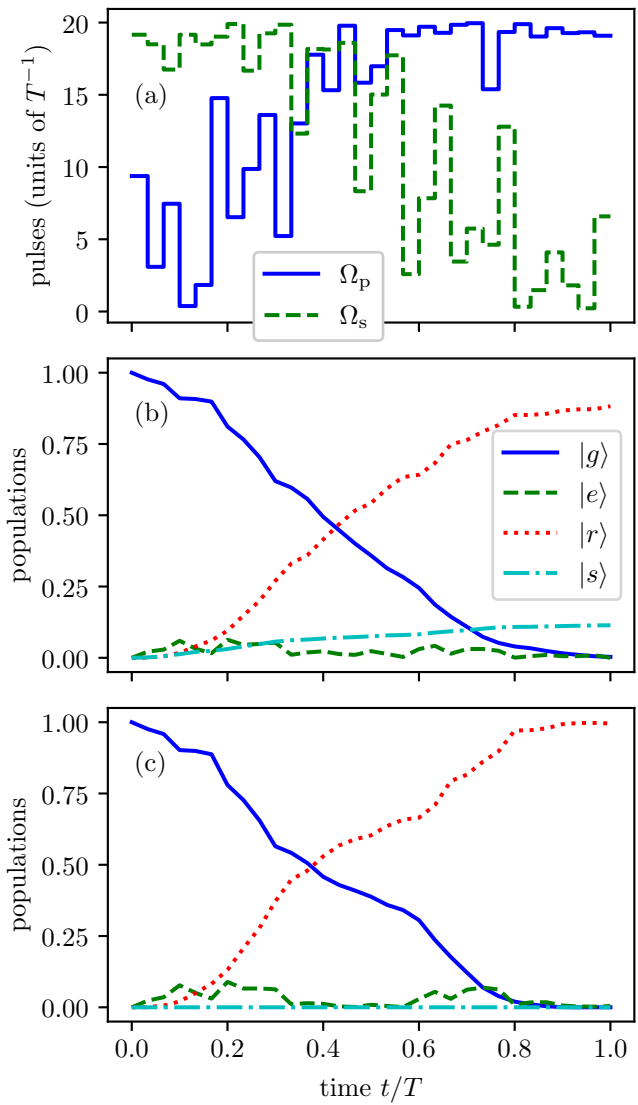}
 \caption{Pulses found by the agent (a) and corresponding population transfer
   for $\gamma=5/T$ (b). Subplot (c) reports the population history if we set
   the decay rate $\gamma=0$ with the same pulses as in subplot (a). The
   fidelity in this last case is $\F = 0.996$.}
  \label{Figure:pulsesREINFORCE}
\end{figure}
\begin{figure}[h!]
  \input{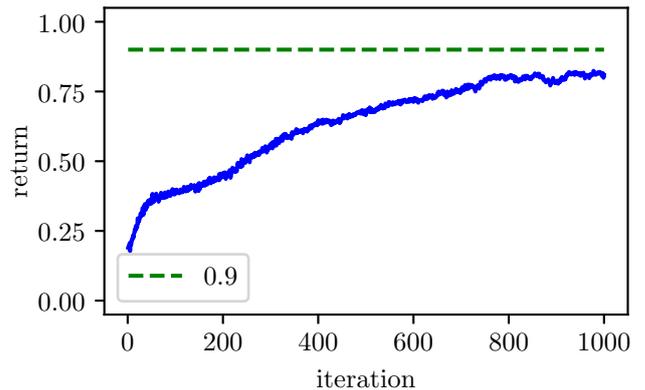}
  \caption{Average final reward of the RL agents in the batch as a function of
    the number of episodes. Here $\gamma=5/T$, $\sigma=0.5$, $batchsize=200$
    and $\eta=0.05$. The Neural Network has two hidden layers of $100$ and $50$
    neurons, respectively, with ReLu activation function [F. Chollet et al.,
    “Keras,” https://keras.io (2015)] while the activation function of the last
    layer is a hyperbolic tangent. The final oscillations are mostly due to the
    choice of the learning rate and they can be shrinked at the cost of a
    bigger number of epochs with a smaller learning
    rate.}\label{Figure:learningREINFORCE}
\end{figure}

\subsection{Policy gradient and REINFORCE}
Giving an exhaustive overview of the various algorithms developed to approach a
generic MDP would be an extremely hard task which goes beyond the scope of this
work. Here we will instead introduce a specific technique and we will directly
apply it to the physical problem introduced in
section~\ref{sec:three-level-popul}. This technique is extremely simple and it
is by no means the state of the art of RL. Nonetheless, we will show that it
allows to address our physical problem.

Recalling the previous section, our final goal is to find the best policy
function $\pi(a|s)$. We can formalize the problem by parametrizing the policy
with a set of parameters $\theta$ so that these parameters can be changed to
find the best policy based on the expected performance. To do this, we can
introduce a performance measure $J(\theta)$ and make use of an approximated
gradient ascent technique to update the parameters $\theta$. Algorithms based
on this approach are referred as policy gradient techniques.

For episodic learning, it can be proven~\cite{Sutton2018reinforcement} that if
we define the performance as the value function starting from the initial state
and following the policy $\pi_{\theta}$, we get the following enstimate for
the gradient of $J(\theta)$
\begin{equation}\label{Epg}
  \nabla J(\theta) = E_\pi [G_t \nabla_\theta\log \pi_{\theta}(A_t|S_t)].
\end{equation}
We hence come up with a stochastic gradient ascent rule for the $\theta$
updates
\begin{equation}\label{pg}
  \theta_{t+1}=\theta_t+\eta G_t \nabla_\theta\log \pi_{\theta}(A_t|S_t),
\end{equation}
where the learning parameter $\eta$ is a real number.

We can then train our agent by (i) initiating an episode following the policy
$\pi_{\theta}$ and taking track of states, actions and rewards, (ii) use
Equation \ref{pg} to update $\theta$ and (iii) repeating (i) and (ii) for
multiple episodes. This algorithm is known as
REINFORCE~\cite{Sutton2018reinforcement}.

There are no contraints on the choice of the function used to parametrize the
policy. However, since we do not have in general prior informations on the
shape of the policy function, the most common choice is to make use of
Artificial Neural Networks due their ability to approximate arbitrarily complex
non-linear function~\cite{HornikNN1989multilayer}, which makes them very
versatile. Moreover, Neural Networks are usually trained via gradient-based
techniques. To be more specific, a cost function $C$ is minimized with respects
to weights and biases of the Neural Network (i.e. its internal parameters) via
stochastic gradient descent or other more advanced gradient based
techniques. Hence, we can implement the policy gradient updates with the right
choice of the cost function.

In general, the Neural Network will take as input a representation of the
state. If, at each step, the agent has to choose over a discrete set of
possible actions (lets say $n$), we can build our Neural Network in such a way
that its output consists in $n$ normalized real numbers that represents the
probabilites for the agent to take one of the $n$ possible actions. The action
will then be randomly chosen with these corresponding probabilities.

However, for many physical problems the action space is continuous. In this
case, rather than parametrizing the policy directly with a Neural Network, we
can assume a specific probability distribution and use a Neural Network to
model some or all of its parameters.

In the following, we will assume a Gaussian policy
\begin{equation}
  \pi_\theta(a|s)=\frac{1}{\sqrt{2\pi}\sigma}e^{-\frac{(a-\mu_\theta(s))^2}{2\sigma^2}},
\end{equation}
where we fix the standard deviation $\sigma$ as an external parameter and we
use a Neural Network to parametrize the mean $\mu(s)$.

\begin{figure}[!ht]
 \centering
 \input{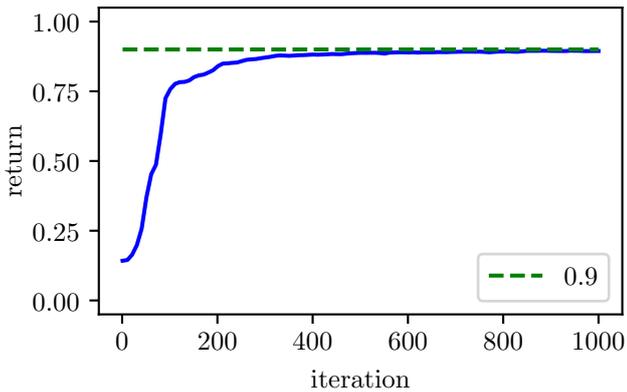}
 \caption{Return as a function of the iteration number for the agent trained
   with TF-Agents~\cite{Guadarrama2018tfagents} for $T\Omega_\mathrm{max}=20$
   and $T\gamma=5$. The neural network used has $3$ hidden layers with $100$,
   $50$ and $30$ neurons, respectively. The activation function is
   \textit{reLU} in each layer except the last one which has a hyperbolic
   secant. The batchsize is $2$ and the \textit{replay buffer} can contain $7$
   episodes.}
 \label{fig:return}
\end{figure}
\begin{figure}[!ht]
 \centering
 \input{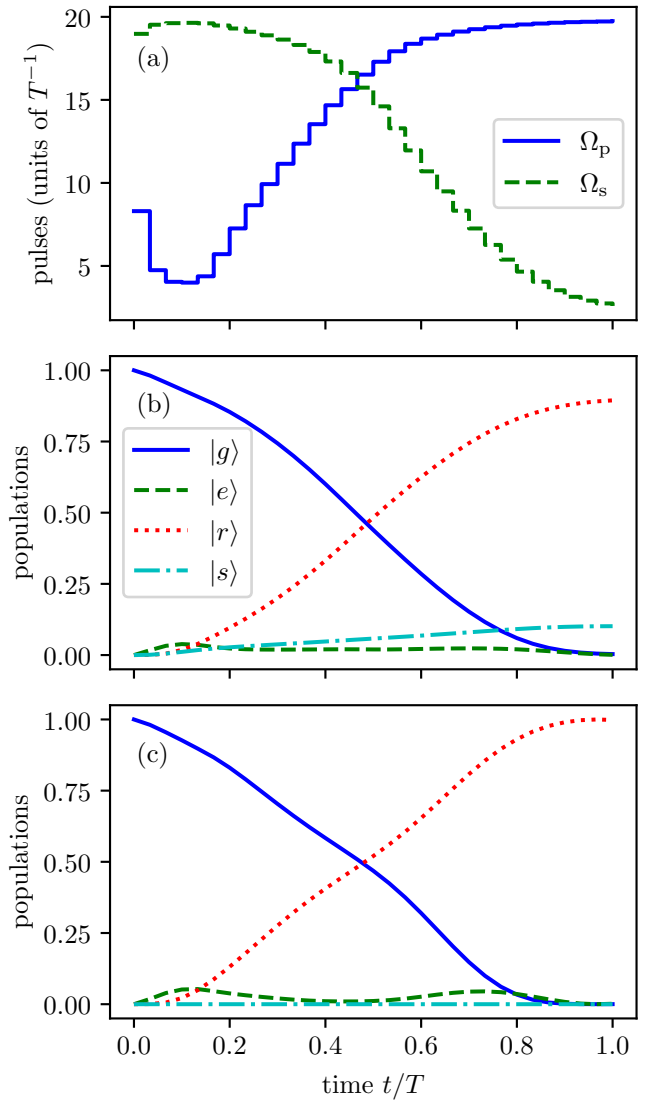}
 \caption{Evolution of the system with the pulses optimized via Reinforcement
   Learning with TF-Agents~\cite{Guadarrama2018tfagents} for
   $T\Omega_\mathrm{max}=20$ and $T\gamma=5$. (a) Pulses and (b) population of
   the states of the system, referring to the last iteration in
   Fig.~\ref{fig:return}. Subplot (c) reports the population history if we set
   the decay rate $\gamma=0$ with the same pulses as in subplot (a). The
   transfer fidelity in this case is $\mathcal{F}>0.998$.}
 \label{fig:example_RL}
\end{figure}

\subsection{Numerical solutions of the RL problem}
Let us now apply the above technique to the physical problem introduced in
Sec.~\ref{sec:three-level-popul}. Our goal is again to find $\Omega_P(t)$,
$\Omega_S(t)$ such that near perfect population transfer from $|g\rangle$ to
$|r\rangle$ for a system evolving according to Equation (2) is achieved during
the interval $[0,T]$.

To formalize the problem, we consider our control terms to be described by
piecewise constant functions (see Sec.~\ref{sec:numer-solut-quant}). We divide
the time interval in $N_{steps}$ smaller intervals $[t_j, t_{j+1}[$ of equal
length. During each of these intervals $\Omega_P(t)$ and $\Omega_S(t)$, take
constant values $\Omega_P(t_j)$, $\Omega_S(t_j)$.

We can now define our MDP. At each step $j$, corresponding to the the time
interval $[t_j, t_{j+1}[$, the agent observation will be given by a
representation of the quantum state of the three-level system plus the sink (i.e $9$
indipendent terms of the density matrix of the system) while the action will give us the
values of $\Omega_P(t_j)$, $\Omega_S(t_j)$.

Specifically, the Neural Network we use to approximate the agent policy will
take as input the 9-dimensional vector
\begin{equation}
  \begin{aligned}
    s_j = \big(&\rho_{gg}(t_j), \rho_{rr}(t_j), \rho_{ee}(t_j), \\
    &\Re(\rho_{ge}(t_j)), \Im(\rho_{ge}(t_j)), \Re(\rho_{gr}(t_j)), \\
    &\Im(\rho_{gr}(t_j)), \Re(\rho_{er}(t_j)), \Im(\rho_{er}(t_j))\big),
  \end{aligned}
\end{equation}
and will give as output two real number
$$\mu_j=(\mu^S(t_j), \mu^P(t_j)) \in [-1,1]^2$$ from which we will sample our
agent actions $$a_j=(a^S(t_j), a^P(t_j))$$ and hence our control terms
\begin{equation}
  \label{eq:OmegaPS}
  \Omega_{P,S}(t_j)=\Omega_0/(1+e^{-3a^{P,S}(t_j)}).
\end{equation}
We define the reward $R_t$ at time step $t$ to be $R_t=0, \forall t=1,N-1$, and
$R_N = \rho_{rr}(T)$. While this is the correct choice in order to represent our
goal (maximizing population on the state $\ket{r}$ at final time $T$), this also
simplifies the REINFORCE algorithm, as we can assign a reward $G_t= R_N =
\rho_{rr}(T) \equiv R$ to all the actions taken by the
agent~\cite{MarquardtSPLN2021machine} in each trajectory. Equation~\ref{Epg} is
then satisfied if we train our Neural Network with stochastic gradient descent
minimizing the cost function
\begin{equation}
  C_j=\frac{1}{2\sigma^2}R|a_j-\mu_{\theta}(s_j)|^2.
\end{equation}
Learning is further enhanced by training the Neural Network in parallel with a
batch of agents (following the same policy $\pi_{\theta}$).

Since we are interested in the best $\Omega_{P,S}(t)$ rather than in the
overall performance of our agents after the training episodes, we continuously
take track of the highest reward reached by the agents and the corresponding
actions.

Numerical results for $\Omega_0 T=20$ are shown in Figure
\ref{Figure:pulsesREINFORCE}. It can be seen that the agents seems to learn
some noisy version of STIRAP-like conterintuitive sequences to achieve
efficient population transfer. In Figure \ref{Figure:learningREINFORCE} we show
the corresponding learning curve by plotting the average final reward of the
agents in the batch for each episode.

We also applied the Reinforce algorithm using the TF-Agents
library~\cite{Guadarrama2018tfagents}. In this case we can easily use some
advanced methods to stabilize and speed up the convergence of the stochastic
gradient descend. In particular we used the Adam
algorithm~\cite{KingmaAC2017adam} and a \textit{replay buffer}. We do not
intend to discuss those methods, but instead just show how they can improve the
learning process, and provide a simple code that can be easily adapted to new
situations. Figure~\ref{fig:return} reports the return as function of the
iteration number, while Fig.~\ref{fig:example_RL} reports the control pulses
and the evolution of the system for the last iteration of
Fig.~\ref{fig:return}. Again the pulses resemble the counter-intuitive pulse
sequence peculiar of STIRAP. Subfig.~\ref{fig:example_RL}(c) reports the
evolution of the system without decay from the intermediate state, but still
drive with the pulses obtained for $T\gamma=5$.

\section{Conclusions\label{sec:conclusions}}
In this tutorial we have introduced the basic concepts of Quantum Optimal
Control and Reinforcement learning. We have shown explicitly how those methods
could be applied to solve a control problem in quantum technology taking as a
reference the process of population transfer in a three-level system, whose one
well-known solution is STIRAP.

A rigorous and thorough comparison between Quantum Optimal Control and
Reinforcement Learning is beyond the scope of this tutorial\footnote{An
  heuristic account has been given in~\cite{ZhangnQI2019when}.}. In fact we did
not use the most advanced or efficient algorithm in either case, for the sake of
keeping the tutorial accessible to a wider audience. However here we highlight
some differences and similarities in our implementations and in our results. The
number of free parameters for QOC is $60$, while for RL is $7644$. The
computational time is also lower for QOC by a factor around $100$, but this
varies greatly in dependence of the available hardware (CPU and/or GPU). Notice
that we also did not optimize the hyperparameters and we suppose that the RL
computational time could improve with a better set of hyperparameters. Both OCT
and RL easily solve the problem by giving STIRAP-like pulses, \ie overlapping
counterintively ordered pulses which tend to occupy the maximal area at their
disposal. The pulses obtained with QOC have a larger area with respect to the
pulses obtained with RL, giving overall a slightly better efficiency. We also
think that with a better set of hyperparaters this difference would be smaller.

Part of the source code developed is open-source and available
online\footnote{\url{https://www.github.com/luigiannelli/threeLS_populationTransfer}}
as a learning tool and can be easily modified to approach similar problems.

\section{Acknowledgements}
The authors thank Alessandro Ferraro, Nicola Macrì, Federico Roy, Phila Rembold,
Riccardo Sessa, Francesco M.~D. Pellegrino, Christiane P. Koch, and Dominique
Sugny for useful discussions. This work was supported by the Northern Ireland
Department for Economy (DfE), the EU H2020 framework through Collaborative
Projects TEQ (Grant Agreement No. 766900), the DfE-SFI Investigator Programme
(Grant No. 15/IA/2864), the Leverhulme Trust Research Project Grant UltraQute
(Grant No. RGP-2018-266), COST Action CA15220, the Royal Society Wolfson
Research Fellowship scheme (RSWF/R3/183013) and International Mobility
Programme, the UK EPSRC (Grant No. EP/T028106/1), the Finnish Center of
Excellence in Quantum Technology QTF (projects 312296, 336810) of the Academy of
Finland, and RADDESS programme (project 328193), Grant No. FQXi-IAF19-06
(``Exploring the fundamental limits set by thermodynamics in the quantum
regime'') of the Foundational Questions Institute Fund (FQXi), the QuantERA
grant SiUCs (Grant No. 731473 QuantERA), and by University of Catania, Piano per
la Ricerca 2016--18 - linea di intervento ``Chance'', Piano di Incentivi per la
Ricerca di Ateneo 2020/2022, proposal Q-ICT.


\bibliography{references,references2}

\end{document}